\documentclass[aps,prl,reprint,superscriptaddress,floatfix]{revtex4-1}
\usepackage{color,units,physymb}
\usepackage{amssymb}
\usepackage{amsmath}
\usepackage{psfrag}
\usepackage{topcapt}
\usepackage{hyperref}
\usepackage{float}
\usepackage{braket}

\usepackage[utf8]{inputenc}

\usepackage{ifpdf}

\ifpdf
\usepackage[pdftex]{graphicx}
\else
\usepackage{graphicx}
\fi

\begin{document}
\title{Single-photon Resolved Cross-Kerr Interaction for Autonomous Stabilization of Photon-number States}

\author{E. T. Holland}
\affiliation{Departments of Physics and Applied Physics, Yale University, New Haven, Connecticut 06520, USA}

\author{B. Vlastakis}
\affiliation{Departments of Physics and Applied Physics, Yale University, New Haven, Connecticut 06520, USA}

\author{R. W. Heeres}
\affiliation{Departments of Physics and Applied Physics, Yale University, New Haven, Connecticut 06520, USA}

\author{M. J. Reagor}
\affiliation{Departments of Physics and Applied Physics, Yale University, New Haven, Connecticut 06520, USA}

\author{U. Vool}
\affiliation{Departments of Physics and Applied Physics, Yale University, New Haven, Connecticut 06520, USA}

\author{Z. Leghtas}
\affiliation{Departments of Physics and Applied Physics, Yale University, New Haven, Connecticut 06520, USA}

\author{L. Frunzio}
\affiliation{Departments of Physics and Applied Physics, Yale University, New Haven, Connecticut 06520, USA}

\author{G. Kirchmair}
\affiliation{Departments of Physics and Applied Physics, Yale University, New Haven, Connecticut 06520, USA}
\affiliation{Institute for Quantum Optics and Quantum Information of the Austrian Academy of Sciences, A-6020 Innsbruck, Austria}
\affiliation{Institute for Experimental Physics, University of Innsbruck, A-6020 Innsbruck, Austria}

\author{M. H. Devoret}
\affiliation{Departments of Physics and Applied Physics, Yale University, New Haven, Connecticut 06520, USA}

\author{M. Mirrahimi}
\affiliation{Departments of Physics and Applied Physics, Yale University, New Haven, Connecticut 06520, USA}
\affiliation{INRIA Paris-Rocquencourt, Domaine de Voluceau, B.P. 105, 78153 Le Chesnay Cedex, France}

\author{R. J. Schoelkopf}
\affiliation{Departments of Physics and Applied Physics, Yale University, New Haven, Connecticut 06520, USA}

\date{\today}

\ifpdf
\DeclareGraphicsExtensions{.pdf, .jpg, .tif}
\else
\DeclareGraphicsExtensions{.eps, .jpg}
\fi

\begin{abstract}
	Quantum states can be stabilized in the presence of intrinsic and environmental losses by either applying active feedback conditioned on an ancillary system or through reservoir engineering.  Reservoir engineering maintains a desired quantum state through a combination of drives and designed entropy evacuation.  We propose and implement a quantum reservoir engineering protocol that stabilizes Fock states in a microwave cavity.  This protocol is realized with a circuit quantum electrodynamics platform where a Josephson junction provides direct, nonlinear coupling between two superconducting waveguide cavities.  The nonlinear coupling results in a single photon resolved cross-Kerr effect between the two cavities enabling a photon number dependent coupling to a lossy environment.  The quantum state of the microwave cavity is discussed in terms of a net polarization and is analyzed by a measurement of its steady state Wigner function.
\end{abstract}

\maketitle
An unavoidable adversary in quantum information science is decoherence.  A large scale quantum computer must implement error correction protocols to protect quantum states from decoherence \cite{nielsen2010}.  A first step toward fault tolerant quantum error correction is the stabilization of a particular quantum state in the presence of decoherence \cite{devoret2013}.  One such implementation uses gate based architectures with measurement and feedback \cite{calderbank1996,laflamme1996,steane1997,geremia2006,sayrin2011,riste2012,riste2013} for the correction of quantum errors.  An alternative approach to active quantum systems is quantum-reservoir engineering (QRE) \cite{poyatos1996,diehl2008,kraus2008,muschik2011} which harnesses persistent, intentional coupling to the environment as a resource.  Both cases require entropy removal yet only QRE employs environmental losses as a crucial part of their protocols.  QRE does not require an external feedback with calculation since the Hamiltonian interactions are designed \textit{a priori} to determine the final state avoiding uncertainty induced by the quantum-classical interface.  In addition, QRE is less susceptible to experimental noise \cite{muschik2012} and in some cases thrives in a noisy environment \cite{vollbrecht2011}.   

QRE has been demonstrated in macroscopic atomic ensembles \cite{krauter2011}, trapped atomic systems \cite{lin2013}, and superconducting circuits \cite{murch2012,geerlings2013,shankar2013}.  Circuit quantum electrodynamics (cQED) systems are an attractive platform for QRE due to the experimental freedom to design strong interactions between superconducting qubits and microwave cavities \cite{wallraff2004}.  Interactions between a superconducting transmon qubit and a microwave cavity have demonstrated qubit-photon entanglement \cite{wallraff2004}, creation of quantum oscillator states \cite{hofheinz2008,hofheinz2009}, and quantum non-demolition measurements of an oscillator \cite{johnson2010}.  Investigations using three dimensional waveguide cavities resulted in increased coherence times \cite{paik2011,reagor2013,pop2014} allowing the observation of novel quantum phenomena such as the single photon self-Kerr effect \cite{kirchmair2013}, a protocol that creates arbitrarily large Schr{\"o}dinger cat states \cite{vlastakis2013}, realtime parity monitoring of the decay of cat states \cite{sun2014}, and a protocol that confines the state of a cavity to a quantum manifold \cite{leghtas2015}, yet no demonstration of a nontrivial cavity state QRE protocol exists.

In this Letter, we demonstrate the single photon resolved cross-Kerr effect between two superconducting microwave cavities which is a new regime of cQED.  This nonlinear coupling causes an excitation in one cavity to change the resonance frequency of the other cavity by more than their combined linewidth.    While the state dependent shift between a qubit and a cavity has been previously observed \cite{schuster2007} in this Letter we present the observation of the state dependent shift between two microwave cavities.  In this work a transmon is used to introduce nonlinearities to the cavities and for tomography.  This new regime of cQED is used to realize the first cQED QRE protocol that stabilizes quantum states of a microwave cavity.  Here we demonstrate a protocol that stabilizes a one photon Fock state of a microwave cavity.  Since the storage cavity is restricted to its first two energy levels, this process can be described as a population inversion and as an effective negative temperature.  This protocol could be extended to higher photon states of the microwave cavity by including more CW drives.  The single photon resolved cross-Kerr is necessary for a QRE protocol that stabilizes cat states of an oscillator \cite{roy2015} and may be used as a cavity-cavity entangling operation.

Within a cQED framework, we model our system as two harmonic oscillators coupled to a nonlinear oscillator \cite{devoret1995}.  The most nonlinear oscillator in our circuit is the transmon whose nonlinearity originates from the Josephson junction with an inductance that is nonlinear with respect to the flux across it.  This system is well described by the following Hamiltonian \cite{paik2011,nigg2012,vlastakis2013}
\begin{align}
\nicefrac{\mathbf{H}}{\hbar}&=\omega_{q} \mathbf{a^{\dagger}a}+\omega_{s} \mathbf{b^{\dagger}b}+\omega_{c} \mathbf{c^{\dagger}c} \nonumber \\
 & {}-\frac{A _{q}}{2}\mathbf{a^{\dagger 2}a^2}-\frac{A _{s}}{2}\mathbf{b^{\dagger 2}b^2}-\frac{A _{c}}{2}\mathbf{c^{\dagger 2}c^2}\nonumber \\
 & {}-\chi _{qs}\mathbf{a^{\dagger}ab^{\dagger}b}-\chi _{qc}\mathbf{a^{\dagger}ac^{\dagger}c}-\chi _{sc}\mathbf{b^{\dagger}bc^{\dagger}c}
\end{align}

The subscripts used in the Hamiltonian are `$q$' for the transmon qubit, `$s$' for the storage cavity, and `$c$' for the cooling cavity.  On the first line are dressed angular frequencies denoted by $\omega_i$.  The second line contains self interaction Kerr type terms, called anharmonicities, of the modes denoted by $A _i$.  On the final line are the state dependent shifts, $\chi _{ij}$, between modes.  Since the state dependent shift, to fourth order, is proportional to the geometric mean of the anharmonicities  of the modes, $\chi_{ij}=2\sqrt{A_iA_j}$ \cite{nigg2012}, a strong, dispersive interaction between modes requires an appreciable anharmonicity for each mode.

The experimental setup consists of two microwave cavities \cite{kirchmair2013,vlastakis2013} machined from high purity (99.99\%) etched aluminum \cite{reagor2013}.  Between the two cavities is a trough where the transmon is positioned to couple the cavities.  Couplers are used to set the dissipation rate of the cooling cavity $\kappa_{\textrm{c}}$, while the coupler to the storage cavity is weakly coupled allowing direct excitation of the storage cavity.  The transmon is fabricated \cite{frunzio2005} on high purity c-plane sapphire and is coupled to each cavity by an antenna which enables a strong, dispersive interaction between the qubit and each cavity \cite{kirchmair2013} $\nicefrac{\chi _{qs}}{2\pi}(21.1~\textrm{MHz})>\nicefrac{\kappa_{\textrm{s}}}{2\pi}(65~\textrm{kHz}) \textrm{ and } \nicefrac{\chi _{qc}}{2\pi}(4.9~\textrm{MHz})>\nicefrac{\kappa_{\textrm{c}}}{2\pi}(1.7~\textrm{MHz})$.
\begin{figure}
\centering
\includegraphics[scale=1]{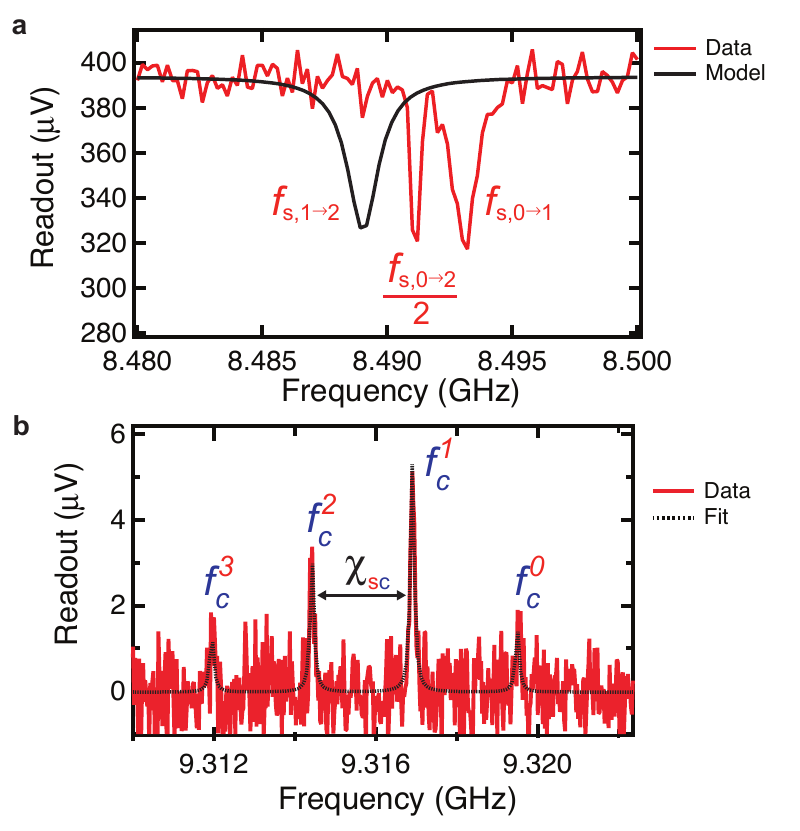}
\caption{Storage and cooling cavity spectra.  
	(a) Spectroscopy is performed on the storage cavity with a single CW drive.  With a large amplitude drive, we observe the two photon transition, $\frac{f_{\textrm{s,}0\rightarrow2}}{2}$.  From this measurement we infer the location of the $f_{\textrm{s,}1\rightarrow2}$ transition (red dashed line) and determine its detuning from the $f_{\textrm{s,}0\rightarrow1}$ transition as 4.0~MHz which we define as the anharmonicity of the storage cavity.
	(b) A 5~ns displacement pulse, whose amplitude gives $~\overline{n}\approx1.5$ in the storage cavity, enables the observation of a single photon resolved cross-Kerr between the two cavities, $\nicefrac{\chi_{sc}}{2\pi}=2.59\pm.06$~MHz.}
{\label{fig:measparms}}
\end{figure}

To measure the storage cavity anharmonicity, we use a single CW drive to perform spectroscopy, Fig.~\ref{fig:measparms}(a).  Using a large amplitude drive which power broadens the $f_{\textrm{s,}0\rightarrow1}$ ($f_i=\nicefrac{\omega_i}{2\pi}$) transition, we observe the two photon transition with frequency $\frac{f_{\textrm{s,}0\rightarrow2}}{2}$.  The detuning corresponds to half the anharmonicity, $A _s$, of the storage cavity and we infer an inherited cavity anharmonicity $\nicefrac{A_s}{2\pi}= 4.0$~MHz.  Following the same method, we determine that the cooling cavity has an anharmonicity $\nicefrac{A_c}{2\pi}=300$~kHz.  To measure the state dependent shift between the two cavities we first perform a 5~ns displacement pulse on the storage cavity, then spectroscopy on the cooling cavity and finally high-power readout \cite{reed2010}.  Shown in Fig.~\ref{fig:measparms}(b) is a spectroscopy measurement of the cooling cavity for a displacement $\overline{n}\approx1.5$ of the storage cavity.  Discrete spectral peaks for up to three photons in the storage cavity are visible. From this we infer a state dependent shift $\nicefrac{\chi _{sc}}{2\pi}=2.59\pm.06$~MHz and observe the first single photon resolved cavity-cavity cross-Kerr.

\begin{figure}
\centering
\includegraphics[scale=1]{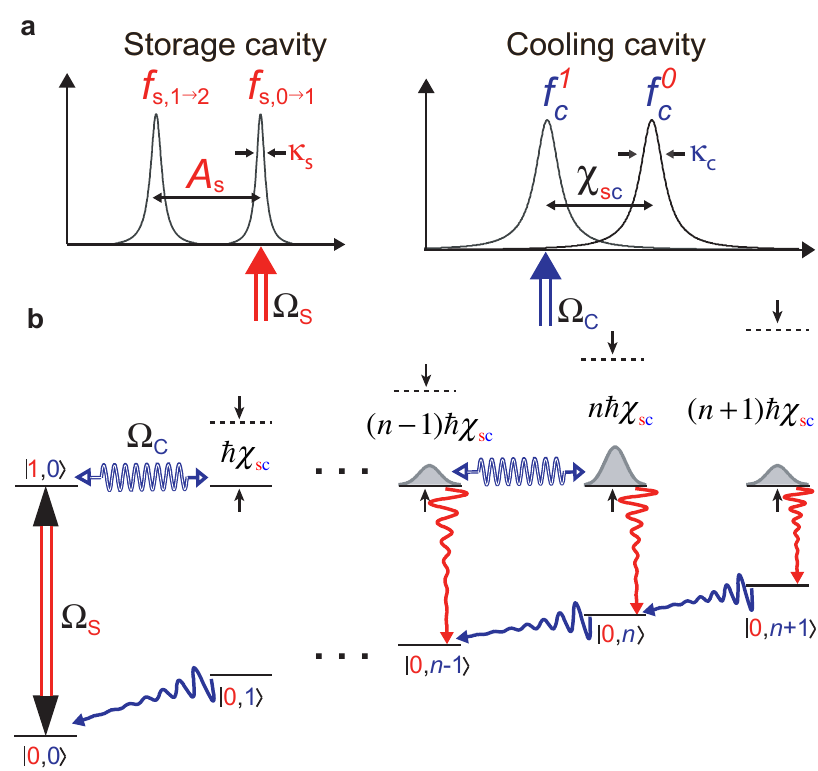}
\caption{Ideal cavity spectrum and Fock state stabilization protocol.
	(a) Left: sketch of idealized storage cavity spectrum.  The storage cavity must have unequal energy levels spacing ($\hbar A_s$), inherited from the coupled qubit, to selectively drive storage cavity transitions.  On the right is the idealized cooling cavity spectrum.  The frequency shift of the cooling cavity due to photons in the storage cavity, the cross-Kerr ($\chi_{sc}$), must be larger than either cavity linewidth to selectively drive this transition. 
	(b) Energy level diagram for the coupled cavity-cavity system tracing over the qubit state.  Ascending vertically are excitations in the storage cavity while to the right is increasing number of excitations in the cooling cavity.  A microwave drive, $\Omega _\textrm{\textbf{S}}$, is applied on the storage cavity so that population only oscillates between vacuum and the first Fock state of the storage cavity.  Simultaneously a drive,  $\Omega _\textrm{\textbf{C}}$, is applied on the cooling cavity such that it is  resonant provided there is exactly one excitation in the storage cavity.  Once resonant, the cooling cavity is pumped to a mean photon number set by the strength of the drive.  The autonomous loop of this protocol is closed by cavity decays, decaying arrows, returning the population to $ \ket{0,0}$ allowing the preparation to be repeated.}
{\label{fig:protocol}}
\end{figure}

The measured Hamiltonian parameters lend themselves well to a cQED QRE protocol that stabilizes Fock states.  The first requirement for this protocol is that the cavity in which Fock states will be stabilized is more anharmonic than its natural linewidth, $A_s>\kappa_s$ so that individual transitions may be selectively driven [Fig.~\ref{fig:protocol}(a) left].  A second requirement is a state dependent shift between the two cavities that is larger than both of their linewidths, $\chi_{sc}>\kappa_s,\kappa_c$ [Fig.~\ref{fig:protocol}(a) right].  In Fig.~\ref{fig:measparms} we see that these requirements are met.  However, this protocol is most successful when the lifetimes of the storage cavity and the cooling cavity are quite different $\kappa_c \gg \kappa_s$.  In Fig.~\ref{fig:measparms}(b) the decay rates of the cavities are comparable, $\left(\nicefrac{\kappa_{\textrm{c}}}{\kappa_{\textrm{s}}}\approx 4 \right)$.  We alter the ratio of lifetimes, while maintaining the necessary requirements $A_s>\kappa_s$ and $\chi_{sc}>\kappa_c \gg \kappa_s$, between the two cavities by increasing the coupling strength to the external environment of the cooling cavity resulting in a new ratio of 25.

Shown in Fig.~\ref{fig:protocol}(b) is a QRE protocol that stabilizes a one photon Fock state in the storage cavity.  This protocol is conceptually similar to the protocol used in Ref. \cite{geerlings2013} which stabilized the ground state of a qubit tensor product with a coherent state of a cavity.  Although we stabilize the ground state of the storage cavity, we also use this protocol to stabilize a one photon Fock state.  Due to the anharmonicity of the storage cavity, a CW drive, $\Omega_{\textrm{S}}$, can be applied to the $f_{\textrm{s,}0\rightarrow1}$ transition.  This drive is an induced Rabi rate between vacuum and a one photon Fock state.  Concurrently, with $\Omega_{\textrm{S}}$, a drive with strength $\Omega_{\textrm{C}}$ is applied detuned by a cross-Kerr from the cooling cavity.  This drive is resonant provided that there is exactly one photon in the storage cavity.  Once resonant, the conditional drive displaces the cooling cavity to a coherent state determined by the amplitude of the drive.  When a photon decays from the storage cavity $\Omega_{\textrm{C}}$, is no longer resonant and the cooling cavity quickly decays to vacuum.  Once back to the ground state, the storage cavity is resonant with the drive $\Omega_{\textrm{S}}$.  This protocol reaches its steady state solution in a time governed by the decay rate of the cooling cavity.  The steady state population in the one photon Fock state of the storage cavity will be determined by its decay rate, $\kappa _s$, and the stabilization rate, $\kappa _{ \uparrow}$.  The stabilization rate is defined as the rate at which the system is returned to the target state when a photon decays from the storage cavity.  Using a simple four state model we expect that to achieve a 99\% one photon Fock state in the storage cavity, a minimum ratio of lifetimes between the two cavities of 300 is required see supplementary material.

This protocol requires both the frequencies of the two microwave drives and their amplitudes be chosen appropriately.  From a full simulation of the Linblad master equation as well as our experimental observations, we find optimal performance when $\Omega_{\textrm{S}}\approx \kappa _{\textrm{c}}$.  We determine the drive power applied to the cooling cavity through a power dependent dephasing measurement  of the transmon qubit applied roughly at one cross-Kerr detuned from the cooling cavity \cite{hatridge2013}.

The experimental implementation begins with CW drives applied simultaneously to the storage and cooling cavity for a duration of $200\kappa^{-1}_{\textrm{c}}$ which is twenty times longer than the time necessary to reach steady state \cite{geerlings2013}.  To measure the photon population in the storage cavity, we stop the drives, wait for photons to decay from the cooling cavity, and apply conditional qubit $\pi$ pulses to determine the photon number in the storage cavity \cite{johnson2010,kirchmair2013}.  We vary both the drive strength and frequency applied to the cooling cavity while maintaining optimal parameters for the storage cavity.

\begin{figure}
\centering
\includegraphics[scale=1]{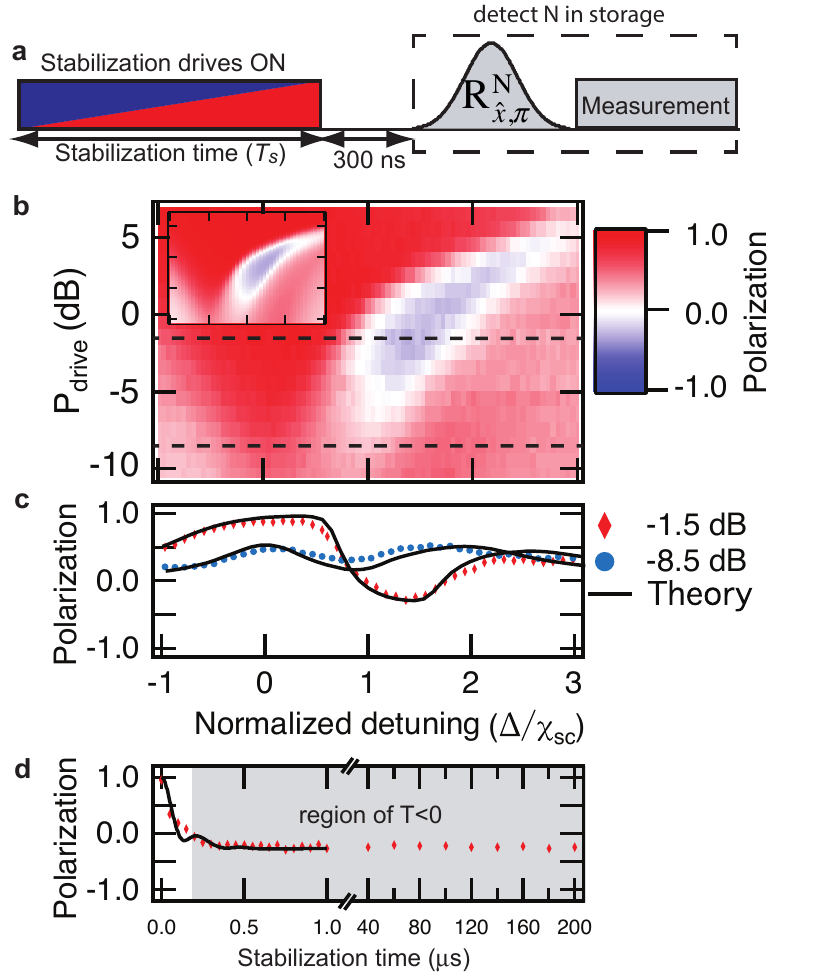}
\caption{Storage cavity polarization.
	(a) The Fock state stabilization protocol described in Fig.~\ref{fig:protocol}(b) is applied for a duration, $T_s$, followed by a 300~ns wait to evacuate excitations from the cooling cavity, a photon selective $\pi$ pulse is then performed on the qubit determining the probability of each photon state of the storage cavity up to three photons.
	(b) Storage cavity state polarization as a function of drive amplitude and frequency.  The frequency of the cooling cavity drive is plotted as $\Delta=f_{\textrm{C}}^0-f_{\textrm{d}}$, and normalized by the cross-Kerr, $\chi_{\textrm{sc}}$.  As the frequency of the drive applied to the cooling cavity is brought in resonance with the first photon peak of the storage cavity $\frac{\Delta}{\chi _{\textrm{sc}}}\approx 1$ the protocol stabilizes the first Fock state of the storage cavity.  The inset is a simulation plot with the same axis and color scale as the experimental result.
	(c) Linecuts for a weak drive power and a drive power resulting in a polarization inversion.
	(d) As the duration of the stabilization protocol is varied the polarization of the storage cavity alters and for infinite time reaches its steady state solution.}
{\label{fig:polarization}}
\end{figure}

We plot the steady state polarization, $p=\frac{P(0)-P(1)}{P(0)+P(1)}$, of the storage cavity after running the protocol in Fig.~\ref{fig:polarization}(b).  $P(n)$ corresponds to the probability of having $n$ photons in the storage cavity.  Due to the selectivity of the drive, $\Omega_{\textrm{S}}$, the storage cavity is limited to its first two Fock states.  We confirm this by measuring populations for the two and three photon Fock states.  When $\Omega_{\textrm{C}}$ is driven at the zero photon peak of the cooling cavity we observe $p=0.95$ demonstrating that storage cavity is overwhelming in the zero photon Fock state despite the induced Rabi drive on the storage cavity.  However, as the drive power and frequency applied to the cooling cavity are varied, steady state stabilization of a polarization inversion occurs corresponding to a predominantly one photon Fock state in the storage cavity.  This population inversion is a purely quantum effect and can be described as an effective negative temperature according to \cite{clerk2010}:
\begin{eqnarray*}
T=\frac{h f_{\textrm{s,}0\rightarrow1}}{2k_{\textrm{B}}\atanh{(p)}}
\end{eqnarray*}
Where $h$ is Planck's constant and $k_{\textrm{B}}$ is Boltzmann's constant.  From this we infer that our steady state solution corresponds to an effective negative temperature of $-0.77\pm0.06$~K in equilibrium with the storage cavity.  

In Fig.~\ref{fig:polarization}(d) we see the time dynamics of this protocol where the initial polarization is unity then changing as a function of time to its steady state value of $p=-0.26\pm0.04$.  Plotted on top of the data is a full simulation of our driven dissipative system where we find excellent agreement in our time dynamics \cite{johansson2013}.  From the four state model, we would expect a polarization of $p=-0.47$.  This value is within a factor of two of both what is measured experimentally and extracted from a full simulation of the Linblad equation.  Through simulation of the full Linblad master equation we find that the limitation in polarization inversion is due to the finite ratio of lifetimes.  Physically realized two cavity systems exist with a ratio of lifetimes being a factor of one thousand \cite{sun2014}.  With that ratio we expect that in steady state $P(1)>0.99$. 
\begin{figure}
\centering
\includegraphics[scale=1]{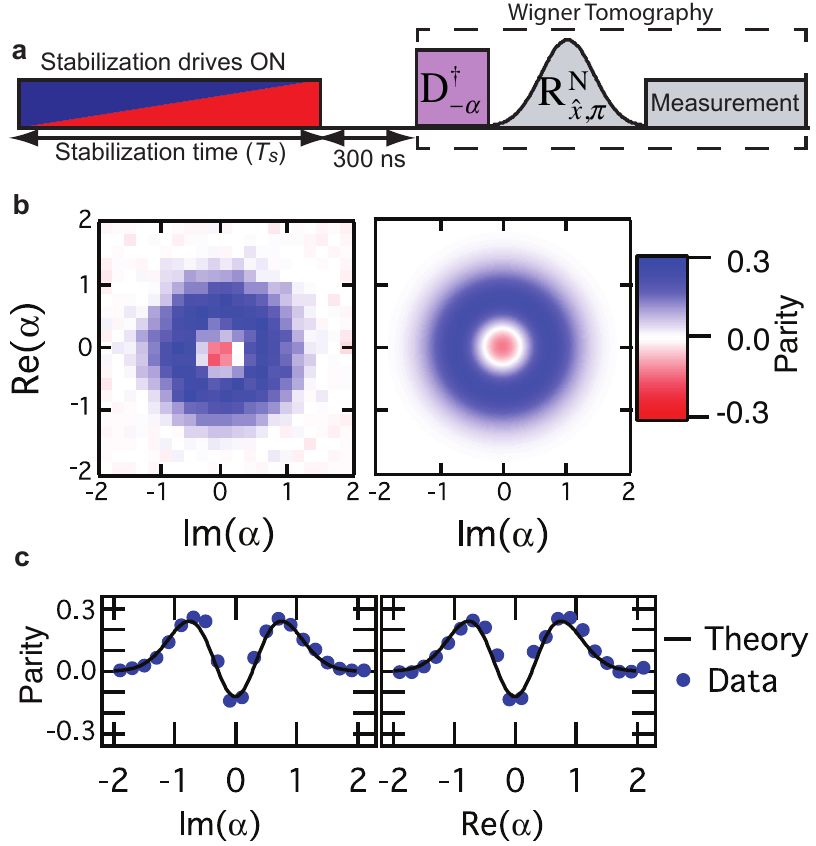}
\caption{Wigner tomography of stabilized steady state of the storage cavity.
	(a) The previously described stabilization protocol is used to reach the desired steady state.  Then Wigner tomography is performed on the state of the storage cavity.
	(b) Left: Measured Wigner function for the steady state of the storage cavity which is a statistical mixture of an $N=1$ and $N=0$ Fock state.  Right: Simulated steady state of the protocol.
	(c) Linecuts along Im$(\alpha)$ and Re$(\alpha)$ for the measured Wigner function and the simulated steady state Wigner function.  Although not a pure $N=1$ Fock state of the storage cavity our long term solution does have negativity in the Wigner function indicative of a quantum state.}
{\label{fig:wigner}}
\end{figure}

Although much of the discussion has framed the storage cavity in the language of spin systems it is still an oscillator.  To demonstrate the oscillator nature of the storage cavity in Fig.~\ref{fig:wigner} we perform cavity tomography measuring generalized Husimi Q functions, $Q_N(\alpha)=\pi ^{-1} \lvert \bra{N}D_{-\alpha}\ket{\Psi}\rvert ^2$ \cite{kirchmair2013}, up to $N=3$  Fock state of the storage cavity, $D_{-\alpha}$ is the displacement operator, and $\Psi$ is the final state and infer the Wigner function by adding and subtracting the even and odd measured Q functions.  We compare these results to the Wigner function of a simulation of the steady state solution to the protocol.  Our results are explained in terms of a harmonic oscillator picture with the steady state of the storage cavity in a statistical mixture of $P(0)=0.37\pm0.03$ and $P(1)=0.63\pm0.02$.  In Fig.~\ref{fig:wigner}~c statistically significant negativity in the Wigner function is observed.  

In conclusion, we present the first single photon resolved cross-Kerr effect between two cavities.  We used this system to implement a cQED QRE protocol that stabilizes Fock states in a superconducting microwave cavity.  We demonstrate one such instance, stabilizing an $N=1$ Fock state, by presenting its reconstructed Wigner function to demonstrate the oscillator nature of the storage cavity.  This protocol can be extended to higher photon numbers of the storage cavity by including more selective microwave drives at the different transitions of the storage cavity.  Our steady state polarization inversion corresponds to $p=-0.26\pm0.04$ which we map to the storage cavity being in equilibrium with a bath of $T=-0.77\pm0.06$~K.  Our protocol is limited by induced spontaneous emission to the environment.  Future implementations would benefit from a Purcell filter and increased nonlinearity in the cQED system.  

This research was supported by the NSF undergrant PHY-1309996, the NSA through ARO Grants No. W911NF-09-1-514 and No. W911NF-14-1-0011, and the IARPA under ARO Contract No. W911NF-09-1-0369.
\bibliographystyle{apsrev4-1}
\bibliography{holland_prl}
\end{document}


\title{Supplemental material for ``Single-photon Resolved Cross-Kerr Interaction for Autonomous Stabilization of Photon-number States"}

\author{E. T. Holland}
\affiliation{Departments of Physics and Applied Physics, Yale University, New Haven, Connecticut 06520, USA}

\author{B. Vlastakis}
\affiliation{Departments of Physics and Applied Physics, Yale University, New Haven, Connecticut 06520, USA}

\author{R. W. Heeres}
\affiliation{Departments of Physics and Applied Physics, Yale University, New Haven, Connecticut 06520, USA}

\author{M. J. Reagor}
\affiliation{Departments of Physics and Applied Physics, Yale University, New Haven, Connecticut 06520, USA}

\author{U. Vool}
\affiliation{Departments of Physics and Applied Physics, Yale University, New Haven, Connecticut 06520, USA}

\author{Z. Leghtas}
\affiliation{Departments of Physics and Applied Physics, Yale University, New Haven, Connecticut 06520, USA}

\author{L. Frunzio}
\affiliation{Departments of Physics and Applied Physics, Yale University, New Haven, Connecticut 06520, USA}

\author{G. Kirchmair}
\affiliation{Departments of Physics and Applied Physics, Yale University, New Haven, Connecticut 06520, USA}
\affiliation{Institute for Quantum Optics and Quantum Information of the Austrian Academy of Sciences, A-6020 Innsbruck, Austria}
\affiliation{Institute for Experimental Physics, University of Innsbruck, A-6020 Innsbruck, Austria}

\author{M. H. Devoret}
\affiliation{Departments of Physics and Applied Physics, Yale University, New Haven, Connecticut 06520, USA}

\author{M. Mirrahimi}
\affiliation{Departments of Physics and Applied Physics, Yale University, New Haven, Connecticut 06520, USA}
\affiliation{INRIA Paris-Rocquencourt, Domaine de Voluceau, B.P. 105, 78153 Le Chesnay Cedex, France}

\author{R. J. Schoelkopf}
\affiliation{Departments of Physics and Applied Physics, Yale University, New Haven, Connecticut 06520, USA}

\date{\today}

\ifpdf
\DeclareGraphicsExtensions{.pdf, .jpg, .tif}
\else
\DeclareGraphicsExtensions{.eps, .jpg}
\fi

\maketitle
\section{System and Hamiltonian Parameters}
The sample is a vertical transmon mounted in a high purity, etched aluminum cavity.  This sample is secured in a dilution refrigerator whose base temperature is 20 mK.  The filtering, magnetic protection and mounting are identical to\cite{vlastakis2013}.  Readout is performed via high power readout of our sample using the high power peak of the cooling cavity\cite{reed2010}.
\begin{figure}[h!]
\centering
\includegraphics[scale=1.5]{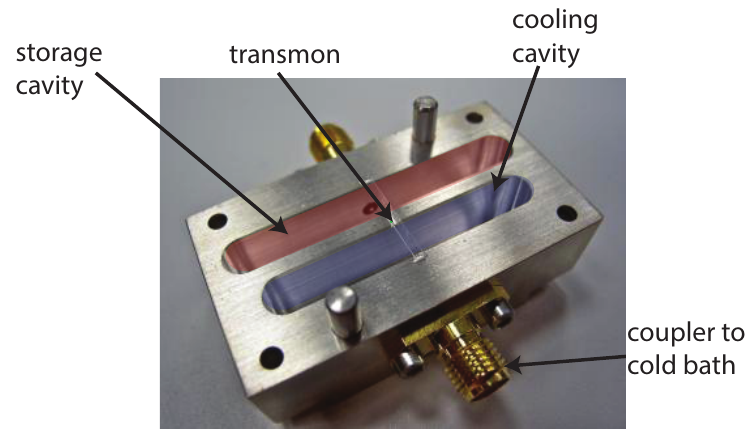}
\caption{Picture of sample}
{\label{fig:sample}}
\end{figure}
Our full Hamiltonian to fourth order in the junction flux is:
{\large
\begin{align}
\nicefrac{\mathbf{H}}{\hbar}&=\omega_{q} \mathbf{a^{\dagger}a}+\omega_{s} \mathbf{b^{\dagger}b}+\omega_{c} \mathbf{c^{\dagger}c} \nonumber \\
 & {}-\frac{A _{q}}{2}\mathbf{a^{\dagger2}a^2}-\frac{A _{s}}{2}\mathbf{b^{\dagger2}b^2}-\frac{A _{c}}{2}\mathbf{c^{\dagger2}c^2}\nonumber \\
 & {}-\chi _{qs}\mathbf{a^{\dagger}ab^{\dagger}b}-\chi _{qc}\mathbf{a^{\dagger}ac^{\dagger}c}-\chi _{sc}\mathbf{b^{\dagger}bc^{\dagger}c}
\end{align}
}
{\large
\begin{center}
  \begin{tabular}{| c | c | c |}
  \hline
    \textbf{Term} & \textbf{Value} & \textbf{Units} \\ \hline
    $\nicefrac{\omega_{q} }{2\pi}$& $7249\pm.013$ & MHz \\ \hline
    $\nicefrac{\omega_{s} }{2\pi}$& $8493\pm.017$ & MHz \\ \hline
    $\nicefrac{\omega_{c} }{2\pi}$& $9320\pm.024$ & MHz \\ \hline
    $\nicefrac{A_{q} }{2\pi}$& $26\pm.3$ & MHz \\ \hline
    $\nicefrac{A_{s} }{2\pi}$& $4.0\pm.1$ & MHz \\ \hline
    $\nicefrac{A_{c} }{2\pi}$& $300\pm79$ & kHz \\ \hline
    $\nicefrac{\chi_{qs} }{2\pi}$& $21.1\pm.05$ & MHz \\ \hline
    $\nicefrac{\chi_{qc} }{2\pi}$& $4.9\pm.04$ & MHz \\ \hline
    $\nicefrac{\chi_{sc} }{2\pi}$& $2.59\pm.06$ & MHz \\ \hline
  \end{tabular}
\end{center}
}

In this section we also include the decay rates of the cavities for the measurements of figure 2 of the main text where $\kappa_s=7.5~$kHz and $\kappa_c=120~$kHz.  The measurements done in figure 4 of the main text for the stabilization protocol are $\kappa_s=65~$kHz and $\kappa_c=1.7~$MHz.

\section{Four Level Model}
To develop intuition as to how well the Fock state stabilization protocol should work we will look at the simplified case of a four state system.  Starting with the state in bottom left as state A we will label states clockwise as B, C, and D.  Our intention is to stabilize state C.  To determine how well we can stabilize this state we must determine the ratio between the rate that state C decays to state D which we will call $\kappa_{\downarrow}$.  The other rate that we must determine is the rate at which our system corrects the C to D decay process and returns the system to state C.  We will call this the stabilization rate and refer to it as $\kappa_{\uparrow}$.  This rate will be a combination of the decay rate from D to A, $\kappa_{DA}$, the rate to be driven from state A to state B, $\Omega_{AB}$, and the rate at which state B is driven to state C, $\kappa_{BC}$.  Since states C and D serve as a proxy of an oscillator we will have the rate at which it decays, $\kappa_{DA}$, being equal to the rate at which it rings up, $\kappa_{BC}$ and just call this rate $\kappa$.  To recap:
\[\kappa_{DA}=\kappa_{BC}=\kappa\]

\begin{figure}[h!]
\centering
\includegraphics[scale=1.5]{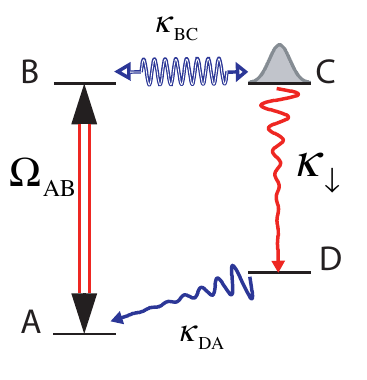}
\caption{Four state system with decay rates}
{\label{fig:fourstate}}
\end{figure}

Now we must determine how $\kappa$ and $\Omega_{AB}$ are used to determine $\kappa_{\uparrow}$.  We identify that if either $\kappa$ or $\Omega_{AB}$ are zero then $\kappa_{\uparrow}$ should also be zero regardless of the value of the other rate.  For this reason we add the rates in inverse:
\[\kappa_{\uparrow}=\left(\frac{1}{\kappa}+\frac{1}{\kappa}+\frac{1}{\Omega_{AB}}\right)^{-1} \]
The optimal choice is for $\Omega_{AB} \approx \kappa$ found through an eigenvalue calculation of the matrix modeling for the four state model.  With this simplification we find:
\[\kappa_{\uparrow}=\frac{\kappa}{3}\]
In the absence of decay from the target state we would expect unit fidelity in creating it.  When accounting for decay we expect that the ratio of $\kappa_{\downarrow}$ and $\kappa_{\uparrow}$ describes how far from unity the stabilized target state will be.  For instance if the two rates are equal we expect that our stabilized state will be produced with a fidelity of 0.5.  In equilibrium we have:
\[\kappa_{\downarrow}P(1)=\kappa_{\uparrow}P(0)\]
Solving for $P(1)$ using that $P(1)+P(0)=1$ we get:
\[P(1)=\frac{1}{1+\frac{\kappa_{\downarrow}}{\kappa_{\uparrow}}}\]
From this we see that if we would like to have a fidelity of 0.99 to the target state the would require a ratio between the decay rate of the cooling cavity, $\kappa$, and the decay rate of the target state, $\kappa_{\downarrow}$ of 300.  From simulation we find that for our actual system that $\kappa_{\uparrow}\approx \frac{\kappa}{9}$.

\section{Linblad Master Equation and Simulation}
We begin with just cavity terms in our Hamiltonian:
{\large
\begin{align}
\nicefrac{\mathbf{H}}{\hbar}&=\omega_{s} \mathbf{b^{\dagger}b}+\omega_{c} \mathbf{c^{\dagger}c}-\frac{A _{s}}{2}\mathbf{b^{\dagger2}b^2}-\frac{A _{c}}{2}\mathbf{c^{\dagger2}c^2}-\chi _{sc}\mathbf{b^{\dagger}bc^{\dagger}c}
\end{align}
}

We add in drives of the form $\Omega_{\textrm{S}}(\mathbf{b^{\dagger}}+\mathbf{b})$ for the storage cavity drive and $\Omega_{\textrm{C}}(\mathbf{c^{\dagger}}+\mathbf{c})$ for the cooling cavity drive.  Where $\Omega_{\textrm{S}}$ and $\Omega_{\textrm{C}}$ are complex drive amplitudes.  By including these drives and entering the rotating frame of both cavities $\mathbf{U_s}=e^{i\mathbf{b^{\dagger}b}\omega_{ds}t}$ and $\mathbf{U_s}=e^{i\mathbf{c^{\dagger}c}\omega_{dc}t}$ we result in detunings of $\Delta_s=\omega_s-\omega_{ds}$ for the storage cavity and $\Delta_c=\omega_c-\omega_{dc}$ for the cooling cavity.  Our driven Hamiltonian is now:
{\large
\begin{align}
\nicefrac{\mathbf{H}}{\hbar}&=\Delta_{s} \mathbf{b^{\dagger}b}+\Delta_{c} \mathbf{c^{\dagger}c}-\frac{A _{s}}{2}\mathbf{b^{\dagger2}b^2}-\frac{A _{c}}{2}\mathbf{c^{\dagger2}c^2}-\chi _{sc}\mathbf{b^{\dagger}bc^{\dagger}c}+\Omega_{\textrm{S}}(\mathbf{b^{\dagger}}+\mathbf{b})+\Omega_{\textrm{C}}(\mathbf{c^{\dagger}}+\mathbf{c})
\end{align}
}
With this driven Hamiltonian we can write the master equation for our system as:
\[\dot{\rho}=-i[\mathbf{H},\rho]+\kappa_sD[\mathbf{b}]\rho+\kappa_cD[\mathbf{c}]\rho\]

The above master equation is used in QuTIP 2.2 \cite{johansson2013} with the steady state solver and the time dependent solver to produce the simulation results that appear in Figure 3 and Figure 4 of the paper.

\bibliographystyle{apsrev4-1}
\bibliography{holland_prl}